# Direct Light Orbital Angular Momentum Detection in Mid-Infrared based on Type-II Weyl Semimetal TaIrTe$_4$

Jiawei Lai,[1] Junchao Ma,[1] Zipu Fan,[1] Xiaoming Song,[2] Peng Yu,[3] Zheng Liu,[4] Pei Zhang,[5] Yi Shi,[6] Jinluo Cheng,[7] Dong Sun, [*,1, 8]

[1]International Center for Quantum Materials, School of Physics, Peking University, Beijing, People's Republic of China

[2]State Key Laboratory of Precision Measurement Technology and Instruments, School of Precision Instruments and Opto-electronics Engineering, Tianjin University, Tianjin, People's Republic of China

[3]State Key Laboratory of Optoelectronic Materials and Technologies, Guangzhou Key Laboratory of Flexible Electronic Materials and Wearable Devices, School of Materials Science and Engineering, Sun Yat-sen University, Guangzhou, Guangdong, China

[4]Centre for Programmed Materials, School of Materials Science and Engineering, Nanyang Technological University, Singapore 639798, Singapore

[5]Shaanxi Province Key Laboratory of Quantum Information and Quantum Optoelectronic Devices, School of Physics, Xi'an Jiaotong University, Xi'an 710049, China

[6]School of Electronic Science and Engineering, Nanjing University, Nanjing, China.

[7]Changchun Institute of Optics, Fine Mechanics and Physics, Chinese Academy of Sciences

[8]Collaborative Innovation Center of Quantum Matter, Beijing 100871, P. R. China

## Abstract

The capability of direct photocurrent detection of orbital angular momentum (OAM) of light has recently been realized with topological Weyl semimetal, but limited to near infrared wavelength range. The extension of direct OAM detection to midinfrared, a wavelength range that plays important role in a vast range of applications, such as autonomous driving, night vision and motion detection, is challenging and has not yet been realized. This is because most studies of photocurrent responses are not sensitive to the phase information and the photo response is usually very poor in the mid-infrared. In this study, we designed a photodetector based on Type-II Weyl semimetal tantalum iridium tellurides with designed electrode geometries for direct detection of the topological charge of OAM through orbital photogalvanic effect. Our results indicate helical phase gradient of light can be distinguished by a current winding around the optical beam axis with a magnitude proportional to its quantized OAM mode number. The topological enhanced response at mid-infrared of $TaIrTe_4$ further help overcome the low responsivity issues and finally render the direct orbital angular momentum detection capability in mid-infrared. Our study enables on-chip integrated OAM detection, and thus OAM sensitive focal plane arrays in mid-infrared. Such capability triggers new route to explore applications of light carrying OAM, especially that it can crucially promote the performance of many mid-infrared imaging related applications, such as intricate target recognition and night vision.

# Main

The applications of optical vortices have made many breakthroughs in rapid succession during the past over 30 years in various field since its birth in 1989,[1] ranging from optical manipulation,[2,3] machining,[4,5] imaging,[6,7] quantum optics[8,9] to optical and quantum communications,[10,11] and even astrophysics.[12,13] The tremendous progress benefits from the rapid developments on the generation and manipulation of optical angular momentum (OAM) of light.[14-16] However, on the detection end, the techniques to measure OAM are usually realized by adopting the interference and diffraction properties, which require indirectly counting the stripes and lattices in the interference or diffraction patterns, or by carrying out indirect phase transformation.[17-19] Direct electric readout of OAM is highly desired which can expand the applications to system-level integration that require on-chip integration and focal plane array imaging with OAM sensitivity.

Recent development on direct characterization of the topological charge of OAM based on type-II Weyl semimetal tungsten ditelluride ($WTe_2$) has opened the prologue of direct electric readout of OAM and subsequent on-chip integration.[20] The detection is achieved at near-infrared wavelength (~1 µm) based on orbital photogalvanic effect (OPGE) driven by the helical phase gradient of light. The OAM mode number can be distinguished by a current winding around the optical beam axis with a quantized magnitude. However, the extension of direct OAM detection to mid-infrared wavelength is technically challenging but highly desired for a vast range of OAM related critical applications, especially for on-chip focal plane array integration for high performance mid-infrared imaging, which are crucial for autonomous driving, night vision and motion detection. The major technical obstacles to realize direct mid-infrared OAM detection are twofold. First, the variation of vector potential associated with OAM of light ($\sim 1/\lambda$) is slow compared to the Brillouin zone size of detection material. Such variation becomes even slower in the mid-infrared compared to that in the near-infrared, which further limits its influence on microscopic processes in material.[21] Second, the absolute photocurrent response is usually poor in the mid-infrared, especially at room temperature, which cannot support enough signal to noise ratio to distinguish the quantized magnitude that is associated with the OAM mode number even such quantization is allowed by the symmetry of the detection material.

Recent development on the photodetection based on topological semimetals has opened possibilities to address the above issues.[22-34] The photocurrent response can be boosted by the large Berry curvature at the vicinity of the Weyl nodes of a Weyl semimetal. When the doping levels match the transition wavelength, such topological enhanced response of Weyl semimetal provides an efficient approach to obtain high responsivity at mid-infrared.[22,35-37] On the other hand, some Weyl semimetals, such as tantalum iridium tellurides (TaIrTe$_4$) share the same crystal symmetry with WTe$_2$,[38-40] which allows quantized OPGE as function of OAM order. When combined with topologically enhance responsivity, it potentially renders the direct orbital angular momentum detection capability in mid-infrared. In this work, we demonstrate that a photodetector based on Type-II Weyl semimetal TaIrTe$_4$ with designed electrode geometries can indeed realize direct detection of the topological charge of OAM through OPGE at 4 µm, a typical mid-infrared wavelength that is widely used for various applications. Our experimental result shows that the direction and amplitude of the photocurrent (PC) driven by OPGE is proportional to the OAM order of the incident beam. The observed OPGE response emerges from the phase gradient of optical fields, rather than other PC generation mechanisms. We also perform the beam-size and spatial dependence measurements to confirm that OPGE current is collected effectively by designed electrode geometries when the beam spot size and position match the electrode structure.

Combined with the linear and circular polarization sensitivity and conventional intensity response, such device is potentially capable of full optical parameter characterizations in mid-infrared when carefully designed into a photodetector array.

**OAM Detection and OPGE in TaIrTe$_4$**

First, we analyze the response of TaIrTe$_4$ under illumination of light with OAM. In the cylindrical coordinate, the most general description of OAM beams with Laguerre-Gaussian (LG) modes propagating in the z direction is given by

$$\vec{E}_{p,m}(r,\phi,z,t) = u_{p,|m|}(r,z)e^{im\phi}e^{i(k_z z-\omega t)}\hat{\epsilon} + c.c. \tag{1}$$

where $u_{p,|m|}(r,z)$ is LG beam profile, the specific mathematical form of which is given in supplementary note 1, and $\vec{k} = k_z \hat{z}$ is the wave vector, $\omega$ is frequency. m is the OAM order, which leads to a phase singularity in the center of the beam when m ≠ 0, thus the central light intensity of the beam is zero and the intensity profile appears like a ring. p is called radial quantum number, which affects the radial electric field distribution of LG beam. An intuitive manifestation of p relates to the number of rings in the intensity profiles (number of rings equals to p+1). In this work, only the LG beams with p=0 are considered. $\hat{\epsilon} = \hat{x} + i\sigma\hat{y}$ represents the polarization of the beam, with σ being the optical helicity or spin angular momentum (SAM) of the beam (with -1≤ σ ≤1, and σ =±1 for circular polarizations). c.c. is the complex conjugate. In addition, we define the ring radius at plane z as the distance between the position that reaches its maximum intensity and the center of LG beam. Under this definition, the ring radius can be written as $\sqrt{\frac{|m|}{2}}w(z)$ for the OAM order m, where $w(z)$ is the radius of basic Gaussian beam at plane z when the spiral phase plates (SPP) used to generate the OAM beam is removed. We note this beam radius definition is different from the conventional one but more straightforward for practical measurement[41].

The generation mechanisms of OPGE in materials belonging to $C_{2v}$ point group are described in detail in Ref. 20 with WTe$_2$ as an example. Figure 1a shows the crystal structure of TaIrTe$_4$, which belongs to point group $C_{2v}$ and its OPGE response should behave qualitatively the same with WTe$_2$, which demonstrates OPGE response in near infrared.[20] The nonlinear photocurrent response to OAM of light arises from the gradient of the laser fields with the inclusion of the electric quadrupole and magnetic dipole effects. After a symmetry consideration as given in detail in supplementary note 2, the second-order dc current density generated from OAM of light can be categorized into four terms according to their dependence on SAM ($\sigma$) and OAM (m), which are experimentally tunable parameters:

$$\boldsymbol{j}^{(dc)}(r,\phi,z) = m \cdot \sigma \, \boldsymbol{j}^{(1)}_{|m|}(r,\phi,z) + m \, \boldsymbol{j}^{(2)}_{|m|,|\sigma|}(r,\phi,z) + \sigma \boldsymbol{j}^{(3)}_{|m|}(r,\phi,z) + \boldsymbol{j}^{(4)}_{|m|,|\sigma|}(r,\phi,z) \quad (2)$$

As an example, the distribution of $\boldsymbol{j}^{(1)}_{|m|}$ in TaIrTe$_4$ can be described by two independent parameters $\beta_1, \beta_3$, which are linear combination of nonvanishing planar rank-4 tensor $\beta_{ijkl}$ as defined by Eq. (S6-S9) in supplementary note 2. $\boldsymbol{j}^{(1)}_{|m|}$ can be written as:

$$\boldsymbol{j}^{(1)}_{|m|}(r,\phi,z) = \left[\hat{\boldsymbol{r}} \left(\text{Im}[\beta_1] + \text{Im}[\beta_3]\cos 2\phi\right) + \hat{\boldsymbol{\phi}} \, \text{Im}[\beta_3]\sin 2\phi\right] \frac{|u_{p,|m|}(r,z)|^2}{r} \quad (3)$$

with $\hat{\boldsymbol{r}}$ and $\hat{\boldsymbol{\phi}}$ being the unit vectors along the radial and azimuthal directions, respectively. The specific expressions of $\boldsymbol{j}^{(2,3,4)}$ are given in Eq. (S3-S5) in supplementary note 2. We note the electric profile term ($u_{p,|m|}(r,z)$) of $\boldsymbol{j}^{(1)}_{|m|}(r,\phi,z)$ is dependent on $|m|$. However, in the measurement, the detected current is not the spatial density itself but an integration of the current density over a certain region along specific direction. Considering that the major beam energy is located around the beam radius, the integration of result is approximately proportional to the beam energy (as shown in detail in supplementary note 3), which is kept the same for beam with different OAM orders in the measurement. Therefore we denote the integrations of the terms $j^{(1)}_{|m|}, j^{(2)}_{|m|,|\sigma|}, j^{(3)}_{|m|}$, and $j^{(4)}_{|m|,|\sigma|}$ over the experimental geometry as capital letters $J^{(1)}, J^{(2)}_{|\sigma|}, J^{(3)}$, and $J^{(4)}_{|\sigma|}$, which do not depend on m and the sign of σ after the integration (see detail calculation in supplementary note 3). In total, the contribution from the m · σ $j^{(1)}_{|m|}$ term after the integration, denoted as $m \cdot \sigma J^{(1)}$, is linear with $m$ as determined by the prefactor $m \cdot \sigma$ when the total beam energy and the beam radius are kept the same for OAM order dependent measurement. The terms $m \cdot \sigma J^{(1)}, m J^{(2)}_{|\sigma|}$, which arise from the helical phase gradient in the azimuthal direction, are proportional to the OAM order $m$ and contribute to the OPGE response. For circularly polarized LG beam, both $m \cdot \sigma J^{(1)}$ and $m J^{(2)}_{|\sigma|}$ contribute to the photocurrents, where the $m \cdot \sigma J^{(1)}$ is proportional to the product of SAM (σ) and OAM (m), and it disappears for linear polarized light. The term $m J^{(2)}_{|\sigma|}$ is nonzero for linear polarized OAM light (σ → 0) and it does not contribute to CPGE. The $\sigma J^{(3)}$ term contribute to CPGE, but doesn't contribute to OPGE. The $J^{(4)}_{|\sigma|}$ term doesn't contribute to either CPGE or OPGE. Experimentally, all these four contributions can be extracted by switching the signs of SAM and OAM of LG beam as given by Eq. (S21-S24) in supplementary note 3. After the integration, the current flowing along both radial and azimuthal direction survives for $m \cdot \sigma J^{(1)}$ and $m J^{(2)}_{|\sigma|}$, which are given by Eq. S12 and S13 in supplementary note 3. To detect the radial and azimuthal current, U-shaped, Ω-shaped and starfish-shaped electrodes can be designed to collect PC. Different shapes of electrodes are sensitive to radial (U-shaped) and azimuthal (Ω-shaped and starfish-shaped) current respectively, thus help to identify and characterize photoresponse from the helical phase profile of the OAM beam.

The mechanism of the OPGE can be understood as light transfers its OAM and energy simultaneously to the electrons. Because the optical phase varies in the azimuthal direction, it induces a spatial imbalance of excited carriers, producing a net current. This is similar to the photon drag effect,[42] which transfers linear momentum and energy from photon to the electron. However, upon normal incidence, the photon momentum being along out-of-plane is forbidden in our measurements. In addition, since the LG beam carries an azimuthal phase profile and annular intensity profile, it can generate spatially dispersive photogalvanic effect (s-PGE) current proportional to the local light intensity gradient.[43] Although part of s-PGE is sensitive to both light helicity and $|m|$, denoted as s-CPGE, it does not switch sign when the vortex beam switch from $+|m|$ to $-|m|$, because the local light intensity is preserved when OAM order reverses sign. The s-CPGE contribution is already included in $\sigma J^{(3)}$. Aside from these PC generation mechanisms from second order nonlinear response, a DC electric field assisted third-order nonlinear effect at the vicinity of the electrodes, denoted as $E_{DC}$-CPGE, can also contribute to circular polarization dependent PC, which will be discussed later on in the last session. Otherwise, PC generated by other mechanisms are not able to exhibit any circular polarization dependence or OAM dependence. A detail discussion on other photocurrent generation mechanisms can be found in supplementary note 4.

**OPGE Response of TaIrTe$_4$**

To perform the measurement, the TaIrTe$_4$ flakes are exfoliated from the bulk material and fabricated into devices with U-shaped electrodes to collect the radial current, as shown in Fig. 1b and Fig. S1a; or into Ω-shaped and starfish-shaped electrodes to collect the azimuthal current, as shown in Fig. S3a and Fig. S4a. The typical thicknesses of TaIrTe$_4$ flakes are between 100 nm and 200 nm. Figure 1c shows the schematic for the OPGE measurement on a U-shaped electrode device. When the OAM order of light is switched from m to -m with fix polarization, the radial current response from OPGE switches sign, whose amplitude is proportional to the OAM order (m). Figure 1d-1f illustrate typical scanning PC microscopy (SPCM) images (Fig. 1e and Fig. 1f) of the TaIrTe$_4$ device with U-shaped electrodes together with the in-situ scanning reflection microscopy (Fig.1d) with 633-nm and 4-μm excitations. The spatial resolutions of these excitations are about 3-μm and 10-μm respectively. The radii of inner and outer electrodes are about 12 μm and 20 μm, and the PC responses are mainly from the region between the electrodes and the sample, which include both the TaIrTe$_4$-metal junctions and the inner area that are between the metal electrodes.

The experimental setup to examine the OPGE response from OAM of light is shown in Fig. 2a. The OAM beams are obtained by passing basic Gaussian beam through a series of SPP with designed OAM order for 4-μm. The SPPs convert basic Gaussian beams into LG modes. The vortex beams are focused by a 40X reflection objective to a ring with radius of 16 μm. The center of the vortex beam is fixed at the center of the U-shaped electrode arcs, which are made of two concentric semicircles and two parallel arms as shown in Fig. 2b. The configuration of the electrodes enables a ∼ 180° solid angle collection of radial current. For the measurement results presented in this part, the ring shape OAM beam is moved to the middle of the inner and outer electrodes. The radius and position dependences of the OPGE response are presented separately in the next session.

To separate out the photocurrent response from OAM of light, a quarter wave plate (QWP) after a polarizer is used for tuning the polarization of vortex beam continuously. By rotating the QWP angle ($\hat{\theta}$) by 180°, the polarization of light undergoes polarization states series from linear ($\hat{\theta}$=0°)-left circular ($\hat{\theta}$=45°)-linear ($\hat{\theta}$=90°)-right circular ($\hat{\theta}$=135°)-linear ($\hat{\theta}$=180°). Figure 2c shows the PC response under constant excitation power of 2.5 mW with different OAM orders. In the measurements of different OAM orders, the radius of the ring is kept the same as defined by the U-shaped electrodes. Benefiting from the topological enhancement of shift current response as reported in our previous work,[22] the device shows appreciable PC response at 4 µm. The PC response can be divided into three components with different $\hat{\theta}$-periodicities after Fourier transform: 180°-periodicity component ($J_C$) which accounts for the CPGE response that is related to different circular polarization ($\sigma = \pm 1$); 90°-period component ($J_L$), which accounts for the anisotropic response to different linear polarization direction; and constant component ($J_0$) that is polarization-independent. The result shows that both positive OAM and negative OAM beams give rise to polarization-dependent PC, and $J_C$ from $+|m|$ and $-|m|$ beams have similar amplitudes but opposite signs, and the amplitude of $J_C$ increases monotonically with m, the OAM order.

According to Eq. (2), both $m \cdot \sigma J^{(1)}$ and $\sigma J^{(3)}$ can contribute to the CPGE response, but only $m \cdot \sigma J^{(1)}$ is proportional to the OAM order m and the response should switch sign when the light polarization switches from left circular to right circular. Figure 3a summarize the extracted CPGE component ($J_C$) as function of OAM order (m). It turns out the CPGE response is dominated by $m \cdot \sigma J^{(1)}$, and $J_C$ displays step-like changes with m ranging from −4 to +4. The quantization of $J_C$ with OAM order implies the contribution from $\sigma J^{(3)}$ to CPGE is minor, because the dependence of $\sigma J^{(3)}$ on OAM order (m) is weak. Meanwhile, we can extract anisotropy response component ($J_L$) and polarization independent component ($J_0$). The m dependences of $J_C$, $J_L$ and $J_0$ are plotted in Fig. 3b. $J_C$ is clearly proportional to m, while $J_L$ and $J_0$ do not show clear m dependence, which suggests $J_L$ and $J_0$ are not responsible for the response to OAM of light. We confirm the results by repeating the measurement on another device with U-shaped electrodes as shown in Fig. S1 and

Fig. S2. Furthermore, we also fabricate devices with Ω-shaped and starfish-shaped electrodes to collect the azimuthal current, and similar step-like changes of Jc with m ranging from −4 to +4 are observed as shown in Fig. S3 and Fig. S4. According to these results the OAM order of light can be clearly distinguished by the quantized plateau of CPGE response.

**Beam-size and Spatial Dependent OPGE Measurements**

When the rings are between the inner and outer electrodes, the current collection efficiency is optimized. The CPGE response is dominated by OPGE in the measurement geometry, and the CPGE contribution from $j_{|m|}^{(3)}(r,\phi,z)$ term is negligible. This is no longer valid when the OAM beam does not match the U-shape electrode when the beam size is getting larger or smaller or the beam center is moved off the center. In the below, we perform the beam-size and excitation position dependence measurements on a device with U-shaped electrodes to study the effects of the beam size and position on the quantization of OPGE response with OAM order, which help sort out other responses that are circular polarization dependent and confirm the OPGE response is along radial direction for U-shaped electrode geometry.

Figure 4 shows the beam-size dependence of photocurrent response at 5 different beam radius excited by light with opposite OAM order (m=±4). As the radius of the vortex beam increases gradually, going through the following conditions: smaller than the inner electrode (7.5 μm, Fig. 4a) - just covering the inner electrode (12 μm, Fig. 4b) - being in the middle of the inner and outer electrode (16 μm, Fig. 4c) - just covering the outer electrode (20 μm, Fig. 4d) - being larger than the outer electrode (28 μm, Fig. 4e), the sign and amplitude of $J_C$ at different OAMs show complicate evolution behaviors as function of beam radius (Fig. 4f). When the vortex beam is in the middle of the inner and outer electrode, $J_C$ amplitudes from the vortex beams carrying opposite

OAM have similar amplitudes but opposite signs, which are consistent with the results concluded in the previous section. As the vortex beams are just covering the inner electrode or the outer electrode, $J_C$ flows along the same direction and doesn't change sign when OAM switches from positive to negative. Furthermore, when the ring radius is smaller than the inner electrode or larger than the outer electrode, the amplitude of $J_C$ becomes much smaller.

The beam size dependent results indicate there are significant CPGE components irrelevant to the OAM of light when the beam diameters don't match the electrodes (Fig. 4a,b,d,e). Such CPGE response could come from the $j^{(3)}_{|m|}(r,\phi,z)$ term which includes s-CPGE effect. Besides $j^{(3)}_{|m|}(r,\phi,z)$ term, it can also come from other effects such as a space charge DC field assisted third-order nonlinear CPGE ($E_{DC}$-CPGE) as reported in previous works.[22,44-46] Although a material with the $C_{2v}$ point group does not support any second-order PGE leading to planar currents from a symmetry consideration, CPGE response is allowed for the third-order nonlinear response t by including a space charge DC field $E_{DC}$. The space charge DC field $E_{DC}$ can be due to multiple effects, such as built-in electric field between the semimetal-metal interface or photothermoelectric effect,but mainly limited to the area at the vicinity of the contacts. $E_{DC}$ switches sign for semimetal-contact and contact-semimetal interface, thus the contribution from $E_{DC}$-CPGE also switches sign. When the vortex beam is in the middle of the inner and outer electrode, $E_{DC}$-CPGE from interfaces with the inner and outer electrodes has opposite sign, so they almost cancel each other. However, when the radius is getting larger or smaller, the ring can only cover one electrode, so $E_{DC}$-CPGE from one electrode contributes and its amplitude is independent on the OAM of light. The $E_{DC}$-CPGE contributes equally to the response for both +4 and -4 vortex beam excitations and the sign of $E_{DC}$-CPGE are opposite with each other for 12-μm and 20-μm excitation which covers different electrodes.

Although the $j^{(3)}_{|m|}(r,\phi,z)$ term can contribute to CPGE, it does not switch sign when the vortex beam switch from $+|m|$ to $-|m|$. Therefore, if we plot the $\Delta J_C = J_C$ (OAM+4)-$J_C$ (OAM-4) and fit it by the expression of collected OPGE current as shown in Fig. 4g, we can find that the evolution behavior of $\Delta J_C$ as function of vortex beam radius is similar to that is observed on WTe2 (Fig. S9 of Ref. 20). The fitting of Fig. 4g follows that the photocurrent collected by a pair of concentric arcs of a U-shape electrodes can be written as integral over the inner electrode r1 to the outer electrode r2 over subtended angle θ: $J_{OPGE} \propto \int_0^\theta \int_{r_1}^{r_2} j_r(\phi) r\, dr\, d\phi$. $j_r(\phi)$ is the local OPGE current generated by the vortex beam at position (r, $\phi$) between the electrodes. Since $j_r(\phi) \propto m \frac{1}{r}|E|^2$, $J_{OPGE} \propto m \int_{r_1}^{r_2} |E|^2\, dr$. For each LG mode, there is a geometrical factor determined by the electrode geometry and beam parameters. Qualitatively, when the diameter of the vortex beam is either smaller than the inner electrode (~7.5 µm) or larger than the outer electrode (~28 µm), the integrand is zero between the two electrodes, so the generated current cannot effectively collected by the electrodes and OPGE component reaches minimum. On the contrary, when the ring radius of the vortex beam locate in the middle of U-shaped electrodes (~16 µm), the OPGE response is optimized. When the vortex beam covers one of the electrodes, either the inner (~12 µm) or the outer (~20 µm), only partial OPGE response is collect and the OPGE lies between the previous cases.

The spatial dependence of OPGE response with fixed beam diameter (~16 µm) is shown in Fig. 5. The OPGE response reaches its maximum when the beam center is at the center of the arcs (at position P3). As the beam center moves away from the center along x direction (either to the left or to the right), or moves down along y direction, the OPGE PC magnitude gradually decreases as shown in Fig. 5c and 5d. This is because the U-shaped electrodes provide a good limit on the PC flowing, the amplitude of the collected OPGE is proportional to the solid angle formed by the center of the beam and the electrodes. The solid angle would decrease whatever direction the beam is move off-centered and thus the OPGE response decreases. The spatial dependent results in the two perpendicular directions together provide an evidence that OPGE current is flowing along radial direction, instead of directions determined by the crystal axes. The beam-size and spatial dependence measurements further confirm that OPGE current is flowing along radial direction and collected effectively by the U-shaped electrodes when the beam spot size and position match the electrode structure.

## Conclusions and Perspectives

Our results demonstrate that photodetectors based on type-II WSM TaIrTe$_4$ with special designed electrode structure can directly resolve the topological charge of OAM of light in the mid-infrared wavelength range. The topological charge of a scalar OAM beam can be resolved by the PC plateau driven by OPGE. We note it has already been demonstrated that TaIrTe$_4$ is sensitive to both the polarization direction of linear polarized light due to its anisotropic photocurrent response,[25] and the helicity of light due to the chirality of Weyl cones as a Weyl semimetal.[47] On the basis of these results, TaIrTe$_4$ represents a class of topological semimetals that is suitable for both OAM and polarization sensitive detection with appreciable mid-infrared response as a result of topological enhancement. The mixture of a variety of OAM orders and the polarization state of light can potentially be resolved simultaneously by a small matrix of well-designed electrodes. Once a device geometry is fixed, characterized, and calibrated, a single device matrix can resolve the full optical parameters of light by back-end processing of the photocurrent signals from different array components. Considering the absolute room temperature detectivity is very challenging to improve for mid-infrared photodetectors, it is crucial for various applications that the image resolution and intricate target recognition capability of a focal plane arrays can be further promoted by the polarization[48-50] and OAM sensitivity as discovered in this work. We expect the results in this work would further push the mid-infrared photodetection based on topological semimetals as a disrupt technology alternative to the conventional roadmap.[35]

**Methods**

Material growth and device fabrication

All the elements used in sample growth were stored and acquired in an argon-filled glovebox. $TaIrTe_4$ single crystals were synthesized by solid-state reaction with the help of Te flux. Single crystals were grown with the same method discussed in Ref. 22. $TaIrTe_4$ flakes were mechanically exfoliated from bulk $TaIrTe_4$ crystal and transferred on to 300nm $SiO_2$/Si substrates. The thickness of $TaIrTe_4$ was estimated to be over 100 nm. Standard electron-beam lithography technique was used to pattern electrodes with designed shapes. The electrodes consisting of 10-nm Cr and 300-nm Au were deposited by electron-beam evaporator,

Optical characterizations

Standard scanning photocurrent mappings were performed in ambient conditions using 632.8-nm He-Ne laser and a CW quantum cascade laser sources emitting at 4 μm. The laser beam was focused by a 40X transmissive objective lens (for 632.8 nm) and a 40X reflection objective (for 4 μm), respectively. Either a scanning mirror or a motorized stage were used to scan the light beam on the device. The reflection signal and PC were recorded simultaneously to get the reflection image and PC response.

For OPGE measurement, the vortex beams carrying different OAM are generated by a series of spiral phase plates designed and manufactured for 4-μm laser. Either ZnSe crystal or silicon was used to manufacture the spiral phase plates to obtain the best transmittance at 4 μm with OAM order up to 4. A motorized rotation stage was used to rotate a quarter waveplate to tune the polarization of the vortex beam. We calibrate the wave plate to rule out experimental artifacts in CPGE component, which may affect the validity of our work (supplementary note 9) The vortex beam was focused by a 40X reflection objective and the spot sizes of the vortex beam can be adjusted by changing distance between the objective and the devices. In both the scanning photocurrent and OPGE measurements, the laser beam was modulated with a mechanical chopper (~373Hz), and the short-circuit PC signal was detected with a current pre-amplifier and a lock-in amplifier.

## Data availability

The data that support the plots within this paper and other finding of this study are available from the corresponding author upon reasonable request.

## Author Information

Corresponding Author

Correspondence to Dong Sun

E-mail: sundong@pku.edu.cn

Affiliations

International Center for Quantum Materials, School of Physics, Peking University, Beijing, People's Republic of China

Jiawei Lai, Junchao Ma, Zipu Fan & Dong Sun

State Key Laboratory of Precision Measurement Technology and Instruments, School of Precision Instruments and Opto-electronics Engineering, Tianjin University, Tianjin, People's Republic of China

Xiaoming Song

State Key Laboratory of Optoelectronic Materials and Technologies, School of Materials Science and Engineering, Sun Yat-sen University, Guangzhou, Guangdong, China

Peng Yu

Centre for Programmed Materials, School of Materials Science and Engineering, Nanyang Technological University, Singapore 639798, Singapore

Zheng Liu

Shaanxi Province Key Laboratory of Quantum Information and Quantum Optoelectronic Devices, School of Physics, Xi'an Jiaotong University, Xi'an 710049, China

Pei Zhang

School of Electronic Science and Engineering, Nanjing University, Nanjing, China.

Yi Shi

Changchun Institute of Optics, Fine Mechanics and Physics, Chinese Academy of Sciences

Jinluo Cheng

Collaborative Innovation Center of Quantum Matter, Beijing 100871, P. R. China

Dong Sun

Author Contributions

D.S. conceived the idea and designed the experiments. P.Y. and Z. L. provided the bulk TaIrTe$_4$ materials. J.W.L. fabricated the device and performed the measurements with help from J.C.M. Z.P.F. and X.M.S under the supervision of D.S.. P.Z. contributed to vertex beam generation used in the experiment. Y.S. contributed to the application perspectives of OAM sensitive detection. J.L.C contributes to the theoretical analysis. J.W.L. and D.S. analyzed the experimental results. J.W.L. and D.S. wrote the manuscript, assisted by J.L.C., Y.S., and P.Z. All the authors comment on the manuscript.

**Competing interests**

The authors declare no competing interests.

**Acknowledgements**

This project has been supported by the National Natural Science Foundation of China (NSFC Grant Nos. 12034001), Beijing Nature Science Foundation (JQ19001). J. L. is also supported by China National Postdoctoral Program for Innovative Talent (BX20200015), China Postdoctoral Science Foundation (2021M690231). P. Y. is also supported by the Plan Fostering Project of State Key Laboratory of Optoelectronic Materials and Technologies of Sun Yat-sen University (No. OEMT-2021-PZ-02). We also thank Junliang Jia for his help in drawing schematic diagram.

# Figures

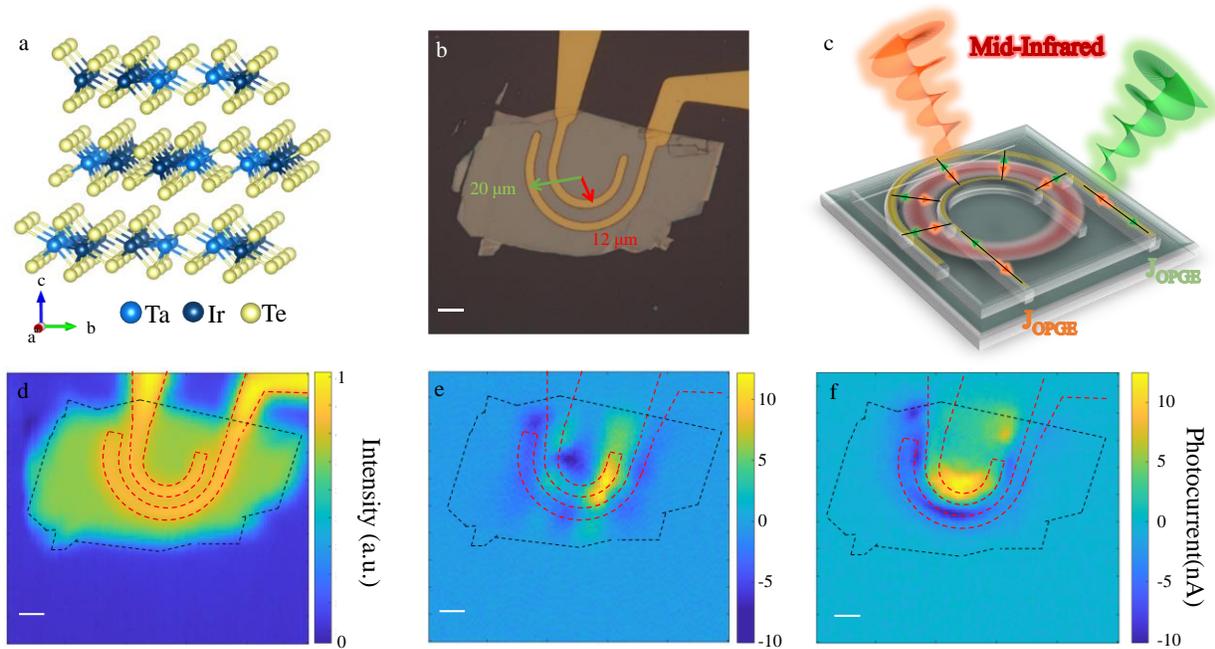

**Figure 1. Basic characterization of the TaIrTe₄ sample.** a. Crystal structures of TaIrTe$_4$ in the $T_d$ phase. b. Optical image of a TaIrTe$_4$ device with U-shaped electrodes. c. Schematic of the OPGE response from light carrying opposite OAM order. d-f. Scanning reflection image (d) and PC microcopic images of TaIrTe$_4$ device with U-shaped electrodes at room temperature with 633-nm (e) and 4-µm (f) excitation, respectively. All scale bars are 10 µm.

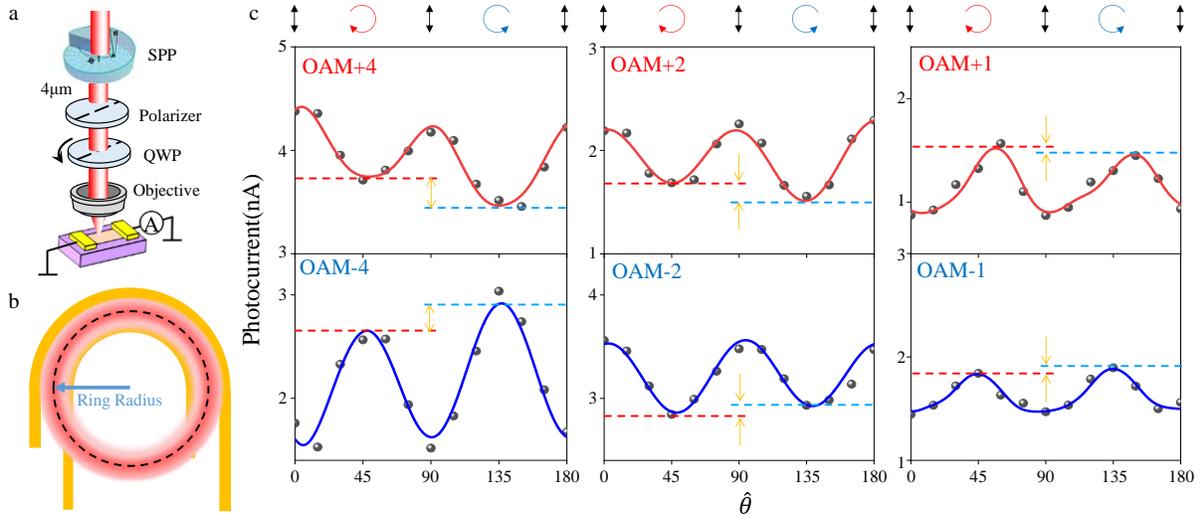

**Figure 2. OPGE PC measurement of TaIrTe$_4$ device with U-shaped electrodes.** a. Schematic diagram of OPGE measurement. b. Schematic diagram of a photodetector device with U-shaped electrodes. The light spot of LG beam is focused at the center of the arc defined by the U-shape electrodes. c. Measured PC amplitudes from OAM +4, +2 and +1(red curve) to -1, -2 and −4 (blue curve) beams, as function of the $\hat{\theta}$. The blue and red dash lines mark the response of right and left circular polarization respectively and the difference between them corresponds to CPGE component of the PC response.

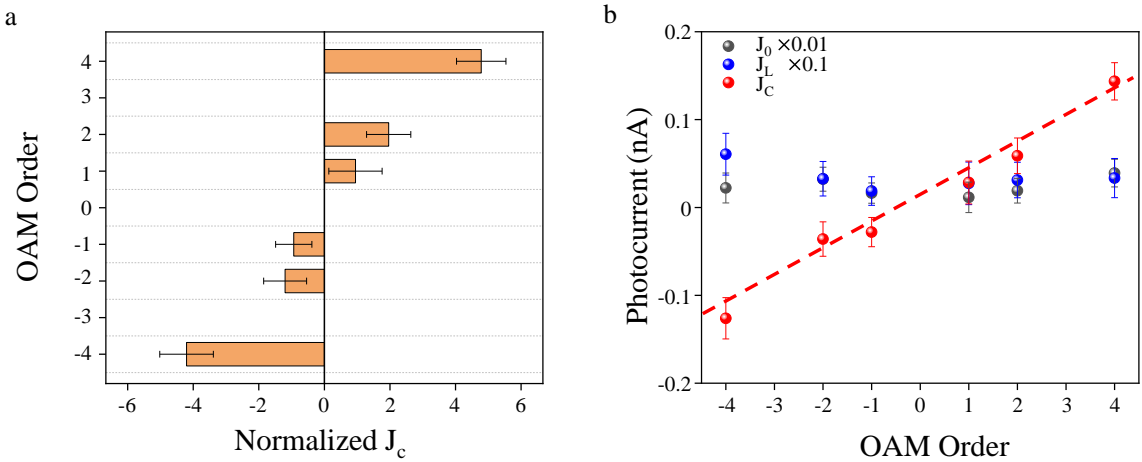

**Figure 3. Evidence of the OPGE PC generated by OAM of light and its dependence on OAM order.** a. Normalized PC that switches with circular polarization, from beams with OAM order ranging from −4 to 4. Error bars represent the standard deviations of the fittings shown in Fig. 2c. b. Three components of the measured PC as function of OAM order: $J_0$, $J_L$, and $J_C$. $J_0$ and $J_L$ are scaled by a factor of 0.01 and 0.1 respectively and their error bars are not scaled to be visible in the plot.

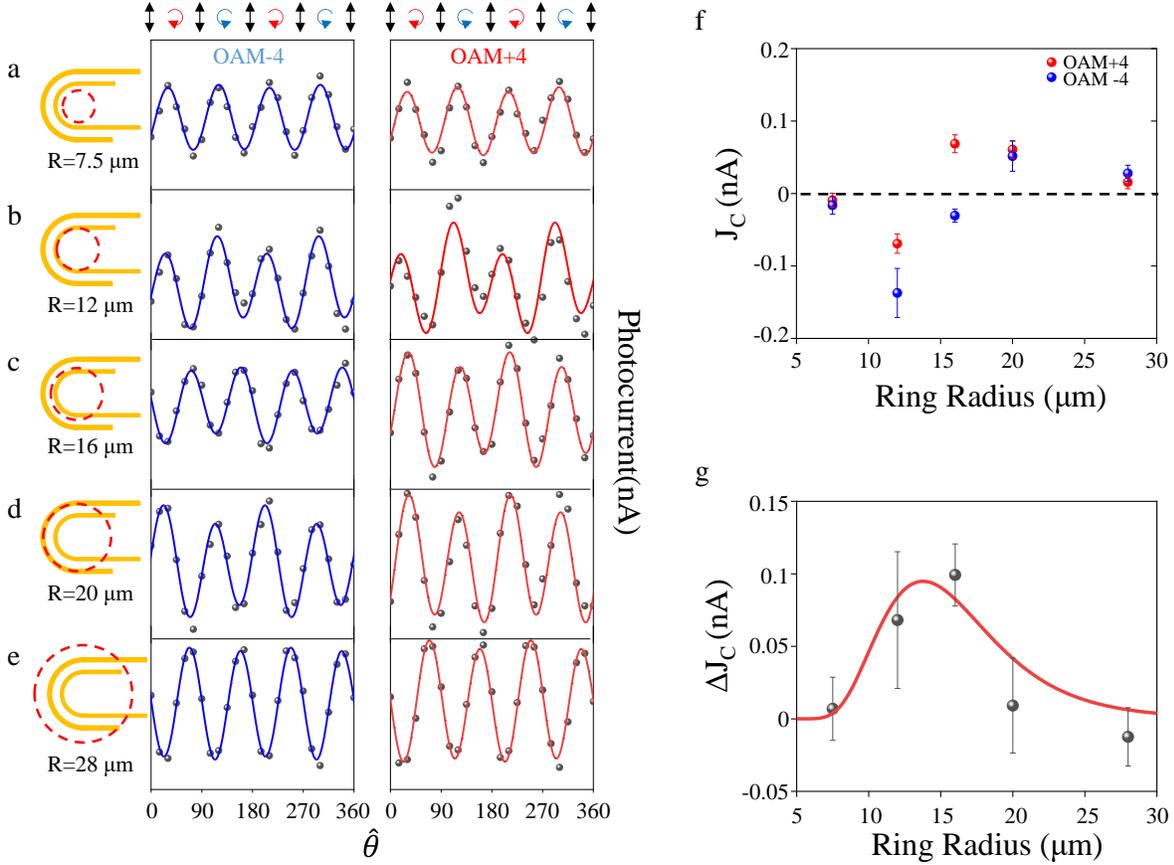

**Figure 4. Spot-size dependent measurements using U-shaped electrodes for OAM +4 and -4 beams.** a-e. PC measured at 5 different spot sizes. All data is plotted in the same scale. The schematic diagrams on the left of each figure show the relative position between the U-shaped electrodes and vortex beam with different spot sizes. f. The $J_C$ component measured at 5 different spot sizes. Error bars represent the standard deviations of the fitting shown in Fig. a-e. g. The $\Delta J_C = J_C(m=+4) - J_C(m=-4)$ as a function of the spot size (black balls) and fitting result (red curve). The fitting parameter is k in $J_{OPGE} = km \int_{r_1}^{r_2} |E|^2 dr$.

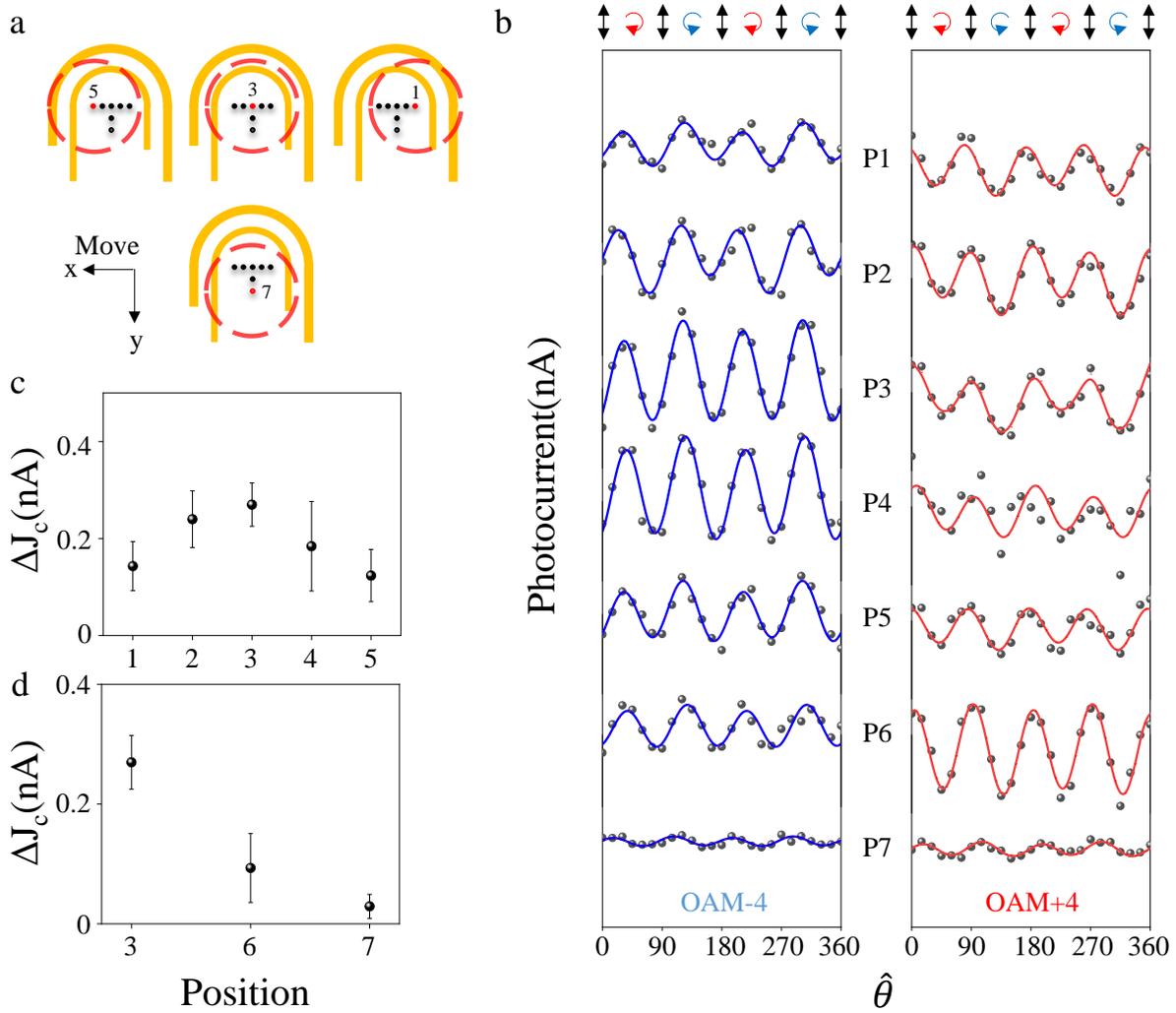

**Figure 5. OPGE PC measurements using U-shaped electrodes while moving the spot position.**
a. Schematic of the beam on a U-shaped electrode device, the beam center locations are marked by the red dots, and the beam profile are marked by the red dash circles. OPGE are measured while moving the beam spot along x and y directions. b. Photocurrent data measured at seven positions, from P1-P7, with $m=\pm 4$. c. The OPGE current magnitude measured at five different positions along x direction. d. The OPGE current magnitude measured at three positions along y direction.